\documentclass[aip,jap,twocolumn,superscriptaddress,longbibliography,10pt]{revtex4-1}

\usepackage{amssymb}
\usepackage{amsmath}
\usepackage{bm}
\usepackage{glossaries}
\usepackage{fancyvrb}

\usepackage{algorithm}
\usepackage{algpseudocode}
\usepackage{silence} 
\usepackage{multirow}
\usepackage{makecell}
\usepackage{graphicx}
\usepackage[usenames,dvipsnames,svgnames]{xcolor}
\usepackage{hyperref}
\hypersetup{
pdfnewwindow=true,          
colorlinks=true,            
linkcolor=Blue,         
citecolor=Blue,         
filecolor=Blue,         
urlcolor=Blue           
}

\newacronym{aimd}{AIMD}{\textit{ab initio} molecular dynamics}
\newacronym{ann}{ANN}{artificial neural network}
\newacronym{dft}{DFT}{density-functional theory}
\newacronym{dp}{DP}{deep potential}
\newacronym{eam}{EAM}{embedded-atom method}
\newacronym{gpu}{GPU}{graphics processing unit}
\newacronym{bpnnp}{BPNNP}{Behler-Parrinello neural-network potential}
\newacronym[longplural={grain boundaries}]{gb}{GB}{grain boundary}
\newacronym{mcmd}{MCMD}{hybrid Monte Carlo and molecular dynamics}
\newacronym{md}{MD}{molecular dynamics}
\newacronym{mc}{MC}{Monte Carlo}
\newacronym{ml}{ML}{machine learning}
\newacronym{mlp}{MLP}{machine-learned potential}
\newacronym{nep}{NEP}{neuroevolution potential}
\newacronym{nn}{NN}{neural network}
\newacronym{rmse}{RMSE}{root-mean-square error}
\newacronym{snap}{SNAP}{spectral neighbor analysis potential}
\newacronym{snes}{SNES}{separable natural evolution strategy}
\newacronym{zbl}{ZBL}{Ziegler-Biersack-Littmark}

\usepackage{siunitx}

\begin{document}

\title{{Solute segregation in polycrystalline aluminum from hybrid Monte Carlo and molecular dynamics simulations with a unified neuroevolution potential}}

\author{Keke Song}
\affiliation{Beijing Advanced Innovation Center for Materials Genome Engineering, University of Science and Technology Beijing, Beijing 100083, China}
\affiliation{Department of Physics, University of Science and Technology Beijing, Beijing 100083, China}

\author{Jiahui Liu}
\affiliation{Beijing Advanced Innovation Center for Materials Genome Engineering, University of Science and Technology Beijing, Beijing 100083, China}
\affiliation{Corrosion and Protection Center, University of Science and Technology Beijing, Beijing 100083, China}

\author{Shunda Chen}
\email{phychensd@gmail.com}
\affiliation{Department of Civil and Environmental Engineering, George Washington University,
Washington, DC 20052, USA}

\author{Zheyong Fan}
\email{brucenju@gmail.com} 
\affiliation{College of Physical Science and Technology, Bohai University, Jinzhou, P. R. China}

\author{Yanjing Su}
\email{yjsu@ustb.edu.cn}
\affiliation{Beijing Advanced Innovation Center for Materials Genome Engineering, University of Science and Technology Beijing, Beijing 100083, China}
\affiliation{Corrosion and Protection Center, University of Science and Technology Beijing, Beijing 100083, China}

\author{Ping Qian}
\email{qianping@ustb.edu.cn}
\affiliation{Beijing Advanced Innovation Center for Materials Genome Engineering, University of Science and Technology Beijing, Beijing 100083, China}
\affiliation{Department of Physics, University of Science and Technology Beijing, Beijing 100083, China}

\date{\today}

\begin{abstract}
One of the most effective methods to enhance the strength of aluminum alloys involves modifying grain boundaries (GBs) through solute segregation. However, the fundamental mechanisms of solute segregation and their impacts on material properties remain elusive. In this study, we implemented highly efficient hybrid Monte Carlo and molecular dynamics  (MCMD) algorithms in the graphics process units molecular dynamics (GPUMD) package. Using this efficient MCMD approach combined with a general-purpose machine-learning-based neuroevolution potential (NEP) for 16 elemental metals and their alloys, we simulated the segregation of 15 solutes in polycrystalline Al. Our results elucidate the segregation behavior and trends of 15 solutes in polycrystalline Al. Additionally, we investigated the impact of solutes on the strength of polycrystalline Al. The mechanisms underlying solute strengthening and embrittlement were analyzed at the atomistic level, revealing the importance of GB cohesion, as well as the nucleation and movement of Shockley dislocations, in determining the material's strength. We anticipate that our developed methods, along with our insights into solute segregation behavior in polycrystalline Al, will be valuable for the design of Al alloys and other multi-component materials, including medium-entropy materials, high-entropy materials, and complex concentrated alloys.
\end{abstract}

\maketitle

\section{Introduction}

Aluminum alloys are widely recognized for their versatility in various structural applications due to their lightweight nature (low density), high strength, and good corrosion resistance~\cite{Schloth2015SM,Wang2007AM}. 
Comprehensively understanding and enhancing the structural properties of aluminum alloys is crucial for unlocking their vast potential in engineering applications. \Glspl{gb} are prevalent defects in metallic alloys that govern the macroscopic strength of materials~\cite{Rollett2004IJMR,Rogers1968SC,King2008SC}. 
Solute segregation at \glspl{gb} profoundly impacts the structure, composition, and overall properties of alloys, offering a promising avenue for developing materials with enhanced performance~\cite{Raade2014COSSMS,Raabe2013AM,Yu2017SC}. 
Despite considerable research on the effects of solute segregation on \glspl{gb} stability and strengthening ~\cite{Lejček2017PMS,Mahjoub2018AM,Wu2016AM,Razumovskiy2015AM,Murdoch2013JMR,Liu2023ASS,WAN2022ASS}, fundamental questions regarding the mechanisms and impacts of solute segregation on material properties persist, primarily due to the challenges associated with observing segregation at the atomic level in situ. 

Atomistic simulation methods, particularly \textit{ab initio} calculations based on quantum-mechanical \gls{dft}, have emerged as valuable tools for understanding solute segregation behavior at \glspl{gb}. For instance, Wu \textit{et al.} investigated the strengthening effect of transition metals on a series of tungsten \glspl{gb}, revealing insights into their dependence on \gls{gb} structure and solute radius \cite{Wu2016AM}. Additionally, Mahjoub \textit{et al.} conducted
extensive \gls{dft} calculations to reveal the general trend of segregation at \glspl{gb} for a large number of solutes in the periodic table~\cite{Mahjoub2018AM}. 
Recently, Song \textit{et al.} studied the effects of the concentration of alloying elements in Fe \glspl{gb} on their strength, hydrogen atom trapping, and hydrogen-induced embrittlement~\cite{Song2024ASS}.

However, \gls{dft} calculations are constrained by efficiency limitations, restricting the analysis of solute atom segregation to specific \glspl{gb} while largely ignoring temperature effects. 
In contrast, \gls{md} simulations with classical potentials allow for the exploration of larger structures while considering temperature effects, providing valuable insights into realistic physical process \cite{Papanikolaou2021ASS,Yuan2017ASS,Huang2023ASS,Koju2020AM}. 
Most previous works have used the \gls{eam} potential or its extensions \cite{Daw1984PRB,Finnis1984PMA}, which often lack the desired accuracy, particularly for alloys.

In this paper, we utilize the highly accurate and efficient machine learning potential, UNEP-v1, recently developed by Song \textit{et al.} \cite{Song2023ARX} as a general-purpose unified \gls{nep} \cite{Fan2021PRB,Fan2022JPCM,fan2022JCP}, for 16 elemental metals and their alloys (Ag, Al, Au, Cr, Cu, Mg, Mo, Ni, Pb, Pd, Pt, Ta, Ti, V, W, Zr), to investigate solute segregation in polycrystalline Al. It has been shown that UNEP-v1 outperforms the \gls{eam} potential by Zhou \textit{et al.} \cite{Zhou2004prb} for various physical properties \cite{Song2023ARX}. To facilitate our investigation, we develop an efficient implementation of the \gls{mcmd} method closely integrated with the \gls{nep} model in the \textsc{gpumd} package \cite{Fan2017CPC}. Employing this method, we systematically study the distribution of 15 solutes in polycrystalline Al. Our analysis reveals that Ag, Au, Cu, Mg, Pb, Pd, and Pt in polycrystalline Al tend to fully segregate at \glspl{gb}, while Ni, Ta, and Zr exhibit partially segregation at \glspl{gb}. Additionally, we observe that Cr, Mo, V, W, and Ti do not segregate at \glspl{gb}, with Ti precipitated as BCC TiAl within the crystal lattice. Subsequently, we investigate the impact of solute segregation on the strength of polycrystalline Al. Our findings indicate that Pd and Pt significantly enhance the strength of polycrystalline Al. Furthermore, we analyze the mechanisms underlying the strengthening and embrittlement associated with solute segregation in polycrystalline Al at the atomistic level, shedding light on the crucial role of grain boundary cohesion, as well as the nucleation and movement of Shockley dislocations, in determining the strength of polycrystalline Al.

\section{Computational details and structural models}

\subsection{The unified neuroevolution potential for 16 elemental metals and their alloys}

We employ the UNEP-v1\cite{Song2023ARX} machine learning potential model based on the \gls{nep}\cite{Fan2021PRB,Fan2022JPCM,fan2022JCP} approach as implemented in the \textsc{gpumd} package\cite{Fan2017CPC}. UNEP-v1 serves as a general-purpose potential for 16 elemental metals (Ag, Al, Au, Cr, Cu, Mg, Mo, Ni, Pb, Pd, Pt, Ta, Ti, V, W, Zr) and their alloys. As demonstrated by Song \textit{et al.} \cite{Song2023ARX}, it  outperforms the \gls{eam} potential by Zhou \textit{et al.} \cite{Zhou2004prb} across various physical properties such as elastic constants, surface formation energy, vacancy formation energy, melting point, phonon dispersions, while maintaining comparable computational speed. For the first time, the UNEP-v1 machine learning potential enabled 100-million-atom atomistic simulations of alloys with \textit{ab initio} accuracy using only eight A100 GPUs \cite{Song2023ARX}.

\subsection{Building the initial polycrystalline Al model}

\begin{figure}[htp]
\hspace*{-1.2cm}
\includegraphics[width=1.28\linewidth]{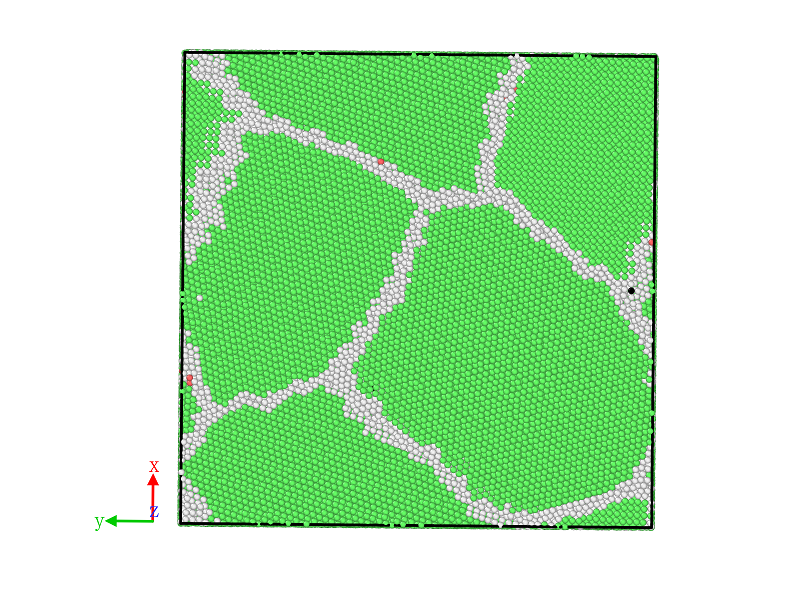}
\caption{
Snapshot of the initial pure polycrystalline Al microstructure obtained from \gls{md} annealing simulation. Green, red, and black spheres represent atoms with local bonding features resembling FCC, HCP, and BCC (very rare) structures, respectively. Gray spheres represent atoms at the \glspl{gb}. 
} 
\label{fig:model}
\end{figure}

We utilized the Voronoi algorithm as implemented in \textsc{atomsk} \cite{Hirel2015CPC} to construct the initial polycrystalline Al model (Fig.~\ref{fig:model}). The model contains 6 grains in a cubic simulation cell, each side of 20 nm, totalling \num{490000} atoms. To release the atomic stress in the \gls{gb} regions, a series of \gls{md} simulations were conducted using \textsc{gpumd} \cite{Fan2017CPC} with the UNEP-v1 machine learning potential \cite{Song2023ARX}. These simulations included a heating process of 200 ps from 300 to 400 K, an annealing process of 200 ps from 400 to 300 K, and finally, an equilibration process of 100 ps at 300 K. These \gls{md} simulations were performed in the $NpT$ (constant number of atoms, pressure, and temperature) ensemble with a zero isotropic target pressure and an integration time step of 1 fs. To study solute segregation, we randomly substitute $1\%$ Al atoms by the solute atoms, before doing the \gls{mcmd} simulations as described below.

\subsection{An efficient MCMD approach implemented in GPUMD}
\label{section:mcmd-algorithms}

\gls{mcmd} proves to be an effective approach for achieving optimal chemical ordering at specific temperatures. In our study, we use the canonical \gls{mcmd} ($N_iVT$ ensemble) to investigate the segregation of solute in polycrystalline Al, where $N_i$ is the number of atoms for species $i$. This canonical \gls{mc} ensemble can be realized by swapping atom pairs with different species. For each \gls{mc} trial, we randomly pick two atoms of different spices and swap their identities, including masses and velocities. The swap is accepted with the probability 
\begin{equation}
    P = \min\left\{1, \exp\left(-\frac{\Delta U}{ k_{\rm B} T}\right)\right\},
\end{equation}
where $\Delta U$ is the change in potential energy due to the trial swap. In \gls{mcmd}, the \gls{mc} and \gls{md} simulations are executed alternately. Specifically, we conduct 100 \gls{mc} trials (regardless of acceptance) after every 100 \gls{md} steps. 

For each \gls{mc} trial swap, evaluating the change in potential energy, $\Delta U$, can be achieved by at least two approaches. A straightforward approach involves recalculating the potential energy of the entire system after each attempted swap (regardless of whether the swap is accepted) and comparing it with the potential energy before the swap, which can be retrieved from the previous accepted \gls{mc} trial. Usually, forces and stresses are also calculated along with energy calculation in this approach. Consequently, one \gls{mc} trial in this approach nearly matches the computational cost of a \gls{md} step. With an equal ratio for \gls{mc} and \gls{md} simulations, \gls{mcmd} is thus twice as costly as pure \gls{md} for completing the same number of \gls{md} steps. This is the approach adopted by \textsc{lammps} \cite{THOMPSON2022cpc} for a general potential model. 

A more efficient approach is to leverage the locality properties of the potential model and  calculate site energies only for atoms that are affected by the attempted swap. In the worst-case scenario, where the two atoms to be swapped are widely separated, we still only need to calculate the site energy for about $4M$ atoms, where $M$ is the average number of neighbors for one atom. The factor of 4 comes from the two atoms (a factor of 2) before and after the swap (another factor of 2). The computational cost for one \gls{mc} trial in this approach is thus independent of the number of atoms $N$ in the system. In this study, where $N=\num{490000}$ and $M \approx 60$, the computational cost of one \gls{mc} trial is only about $0.05\%$ of that for one \gls{md} step. With an equal ratio for \gls{mc} and \gls{md} simulations, this \gls{mcmd} approach is nearly as fast as pure \gls{md} for completing the same number of \gls{md} steps. In other words, the \gls{mc} part is \textit{nearly free} in this approach when $N\gg M$. This is the approach we implemented in \textsc{gpumd} \cite{Fan2017CPC} during the course of this study.
A sample input script for performing \gls{mcmd} simulation in \textsc{gpumd} is provided in Appendix \ref{sec:run_in_MCMD}.

\begin{figure}[htp]
\centering
\includegraphics[width=0.92\linewidth]{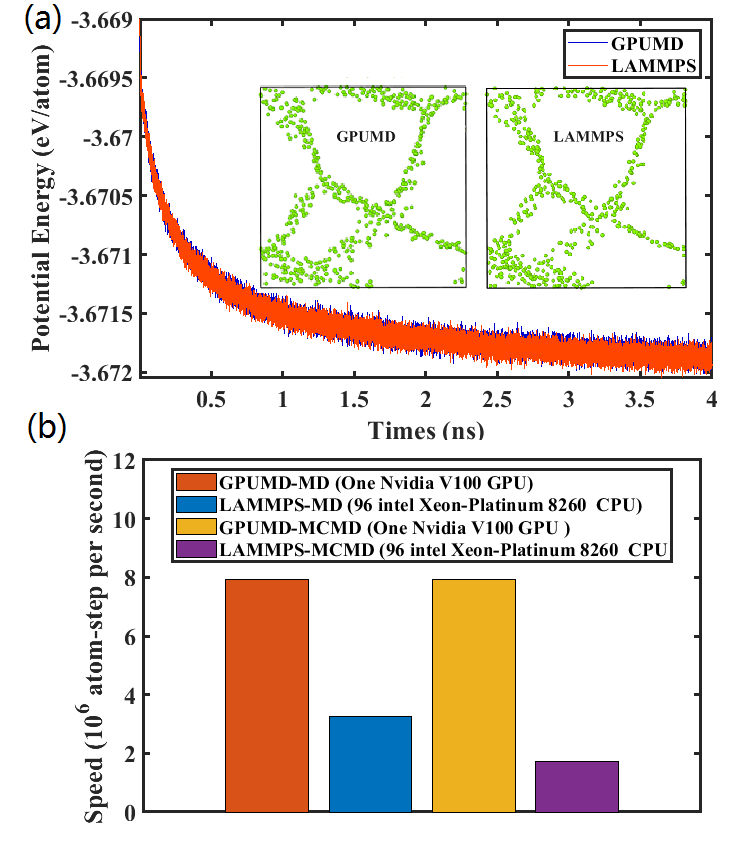}
\caption{(a) Potential energy versus MCMD simulation time for Mg solute segregation in polycrystalline Al, using MCMD methods implemented in GPUMD and LAMMPS. Insets illustrate the distribution of Mg atoms in the 2-nm-thick slices (xy plane, z-axis range 10 to 12 nm) extracted from the equilibrated structures obtained from MCMD simulations. (b) Computational speed comparison between GPUMD (utilizing a single Nvidia Tesla V100 GPU) and LAMMPS (utilizing 96 intel Xeon-Platinum 8260 CPUs). Note: The GPU and CPU resources mentioned above are of comparable price. The MCMD approach implemented in GPUMD demonstrates a 4x speedup compared to LAMMPS. Utilizing a single RTX4090 GPU (which is even lower in price) can further enhance the speedup to about 10x (not shown here), attributed to the efficient GPU implementation and the \textit{nearly cost-free} MC part in the MCMD algorithm in GPUMD.} 
\label{fig:mcmd}
\end{figure}

\subsection{Tensile loading}

During dynamic tension, uniaxial tensile strain was applied at a constant strain rate of $10^8$ s$^{-1}$ by continuously scaling the atomic coordinates and box dimension along the $y$
direction. Throughout the simulation, the $NVT$ ensemble was simulated at 300 K with the integration time step set to 1 fs. Atomic configuration  and dislocation analysis were performed using the Open Visualization Tool\cite{Stukowski2010MSMSE}.
A sample input script for performing tensile loading simulation in \textsc{gpumd} is given in Appendix \ref{sec:run_in_tensile}.

\section{Results and discussion}

\subsection{Comparison of MCMD methods implemented in GPUMD and LAMMPS}

Before applying the \gls{mcmd} method we implemented in \textsc{gpumd} \cite{Fan2017CPC} to systematically study solute segregation in polycrystalline Al, we first evaluate its performance compared to the implementation in \textsc{lammps} \cite{THOMPSON2022cpc}, using Mg solute as an example. Both algorithms produced equivalent physical results, as evidenced by the nearly identical time evolution of the potential energy shown in Fig~\ref{fig:mcmd} (a). Furthermore, the configurations of Mg segregation at the \glspl{gb} are also highly consistent between the two algorithms, as indicated in the insets of Fig~\ref{fig:mcmd} (a).

Regarding the computational efficiency (Fig~\ref{fig:mcmd} (b)), the \gls{nep} model as implemented in \textsc{gpumd} achieves a speed of about $7.9\times 10^6$ atom-step/second in \gls{md} simulations using a single V100 GPU. In contrast, the \gls{nep} model as implemented in \textsc{lammps} reaches a speed of only about $3.3\times 10^6$ atom-step/second, using 96 Intel Xeon Platinum 8260 CPUs. When using \gls{mcmd} with an equal ratio for \gls{mc} and \gls{md} simulations, the \textsc{gpumd} implementation achieves nearly the same speed as in pure \gls{md} simulations, while the \textsc{lammps} implementation experiences an almost halved speed. These timings are expected according to our analyses in Sec.~\ref{section:mcmd-algorithms}. It is noteworthy that the GPU and CPU resources mentioned above are of comparable price. The \gls{mcmd} approach implemented in \textsc{gpumd} demonstrates a 4x speedup compared to \textsc{lammps}. With a single RTX4090 desktop GPU (which is even lower in price) or a single A100 GPU, the speedup is about 10x (not shown here).

\subsection{Segregation of solute in polycrystalline Al}

\begin{figure*}[htp]
\centering
\includegraphics[width=1\linewidth]{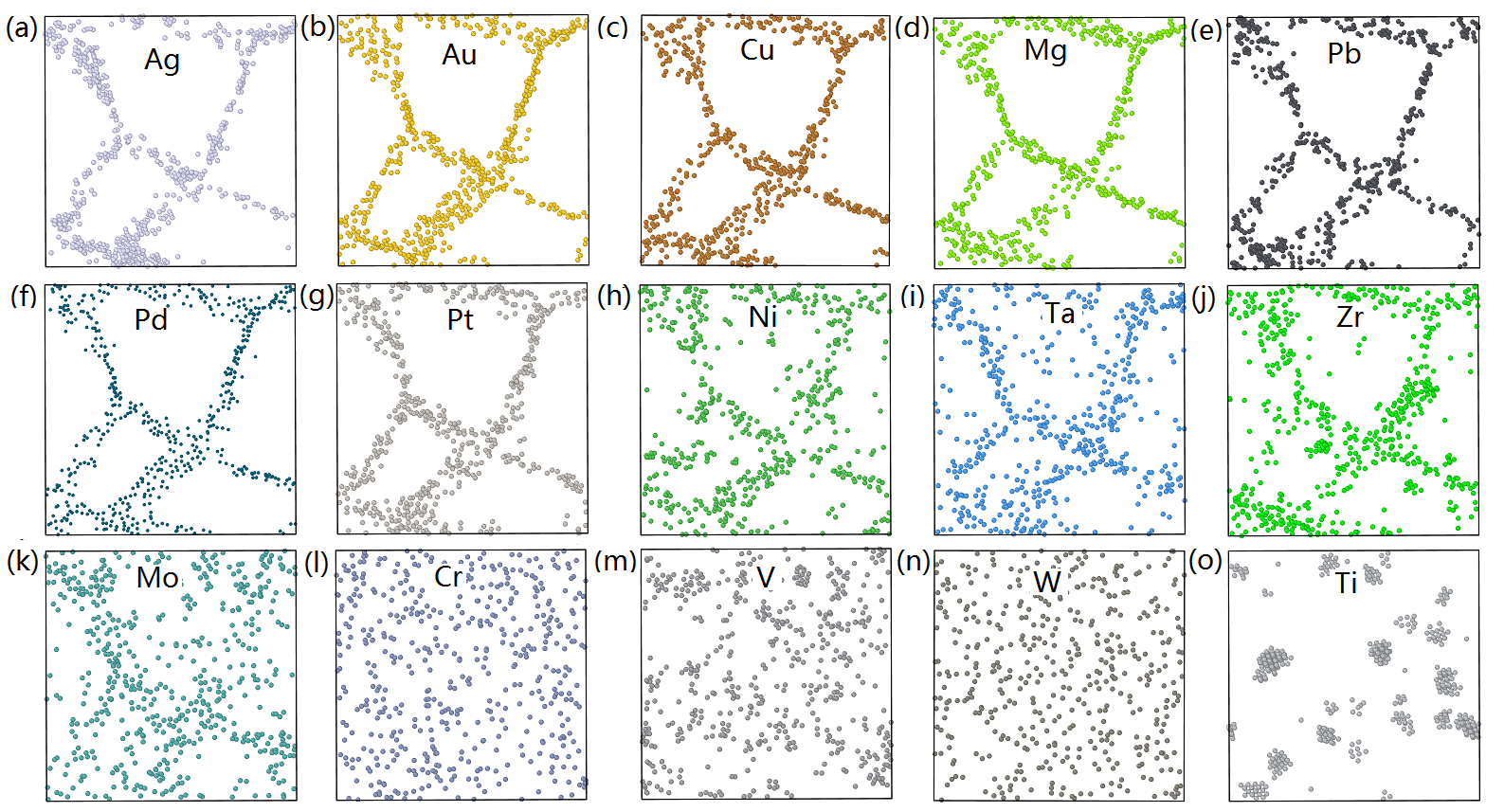}
\caption{Distribution of solute atoms in the 2-nm-thick slices (xy plane, z-axis range 10 to 12 nm)   extracted from the equilibrated structures obtained from MCMD simulations. (a-g) for Ag, Au, Cu, Mg, Pb, Pd, Pt (Category 1: Fully segregated to GBs); (h-k) for Ni, Ta, Zr, Mo (Category 2: Partially segregated to GBs); (l-n) for Cr, V, W (Category 3: Random distribution); and (o) for Ti (Category 4: Precipitation within the grains).} 
\label{fig:segregation}
\end{figure*}

Using the developed \gls{mcmd} algorithm in \textsc{gpumd}, we systematically simulated 15 binary systems where $1\%$ of the Al atoms in the initial polycrystalline Al structure were randomly selected and substituted by one of the following species: Ag, Au, Cr, Cu, Mg, Mo, Ni, Pb, Pd, Pt, Ta, Ti, V, W, and Zr. 
After 5 million \gls{md} and \gls{mc} steps, the potential energies for all binary systems are well converged, and the atomic configurations can be regarded as fully equilibrated. 
Examination of the equilibrium configurations reveals diverse segregation behaviors for the 15 species in polycrystalline Al, which can be broadly categorized into the following four categories:
\begin{itemize}
\item Category 1 (7 species): Ag, Au, Cu, Mg, Pb, Pd, Pt (Figs.~\ref{fig:segregation}(a)-\ref{fig:segregation}(g)). For these species, the solute atoms all segregate to the \glspl{gb}.
\item Category 2 (4 species): Ni, Ta, Zr, Mo (Figs.~\ref{fig:segregation}(h)-\ref{fig:segregation}(k)). For these species, the solute atoms partially segregate to the \glspl{gb} while remaining partially in the grains.
\item Category 3 (3 species): Cr, V, W (Figs.~\ref{fig:segregation}(l)-\ref{fig:segregation}(n)). For these species, the solute atoms are more or less randomly distributed in the system.
\item Category 4 (1 species): Ti (Fig.~\ref{fig:segregation}(o)). In this case, the solute atoms precipitate within the grains.
\end{itemize}

From the above four segregation behaviors, we can see that the segregation concentration of Ag, Au, Cu, Mg, Pb, Pd, and Pt solutes in polycrystalline Al is equal to or greater than the initially set 1\% atomic ratio. Among them, Mg and Cu have also been experimentally\cite{Pickens1987MTA,Jones2001MMTA,Malis1982JMS,SHA2011UL}  and theoretically\cite{Liu1998AM,Liu2005JPCM,Zhang2012APL,ZHAO2018AM} verified. The segregation concentration of Ni, Ta, Mo and Zr is less than the initially set concentration of 1\%. The observation that Ti does not segregate at \glspl{gb} aligns with DFT calculations\cite{Karkina2016CMS}. The precipitation of TiAl within the crystal is plausible, as TiAl metallic compounds are known to be stable and have been extensively studied\cite{Clemens2013AEM,Ozge2022JAC,Wu2006IN}. This is expected to influence the properties of the Al alloy. It is noteworthy that Pb segregation in polycrystalline can cause lattice expansion by $0.4\%$, while other solutes affect the lattice within $0.05\%$. Even if all Pb atoms segregate to the \glspl{gb}, the polycrystalline material will still expand, significantly affecting its strength. In the \gls{gb} region, atomic arrangement differs from the bulk, resulting in the formation of loose and compressed sites due to atomic distortions. These sites offer ample opportunities for the segregation of solutes of varying sizes. 

Given that our model surpasses the size calculated by DFT, the \gls{gb} region provides a complex and more realistic environment that better reflects the solute segregation behavior. This aspect offers valuable insights for future simulations of solute segregation in polycrystalline Al and for the experimental design of solute segregation components in Al alloys.

\begin{figure*}[htp]
\centering
\includegraphics[width=1\linewidth]{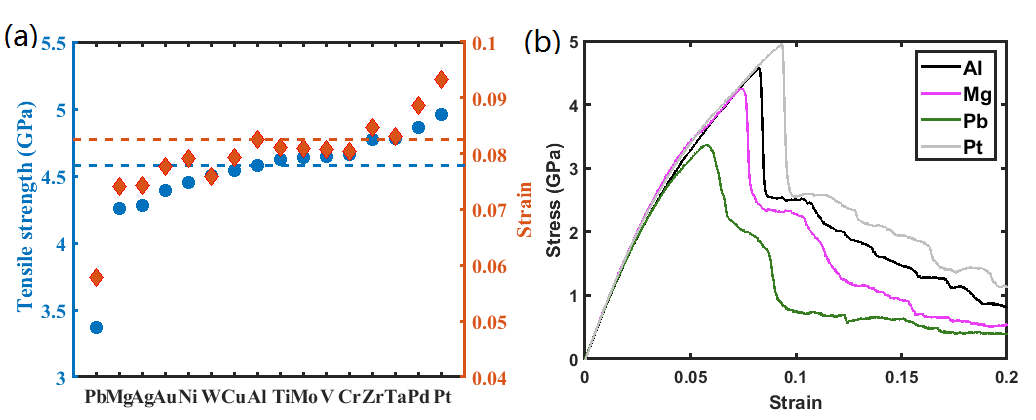}
\caption{(a) Tensile strength as a function of solutes. Blue spheres represent the tensile strength of the polycrystalline binary systems, displayed on the left Y-axis, while red diamonds represent the corresponding strain at the tensile strength of the polycrystalline system, indicated on the right Y-axis. Blue dashed line and red dashed line correspond to values of pure polycrystalline Al, respectively. (b) The stress-strain curves for Al, Mg, Pb, and Pt.} 
\label{fig:stress-strain-curves}
\end{figure*}

\begin{figure}[htp]
\centering
\includegraphics[width=1\linewidth]{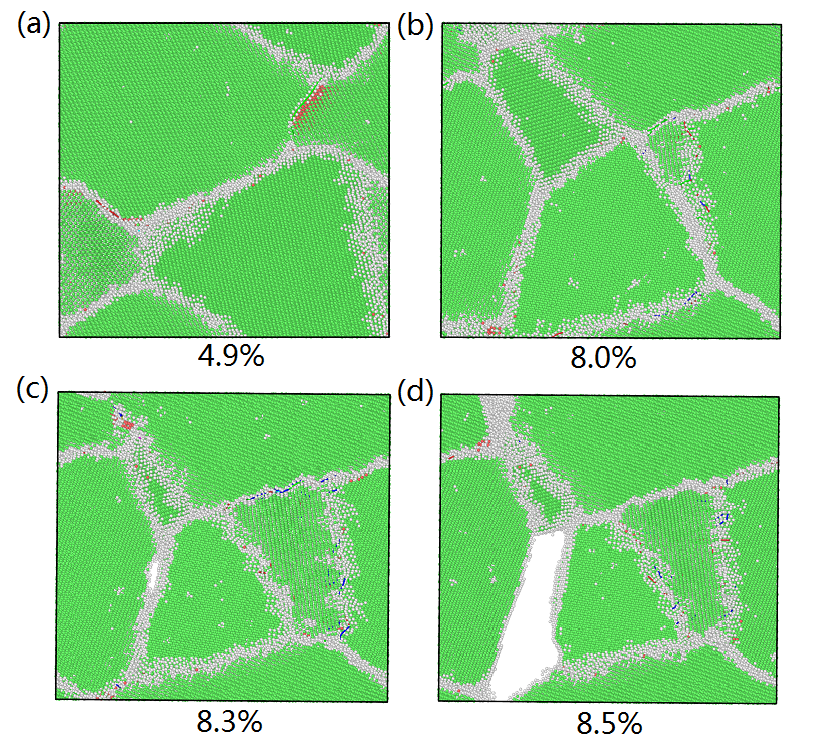}
\caption{Snapshots of 2-nm-thick slices for pure polycrystalline Al under different strains: (a) 4.9\% (yz plane, x-axis range 0.8 to 2.8 nm), (b) 8.0\% (xy plane, z-axis range 11 to 13 nm), (c) 8.3\% (xy plane, z-axis range 11 to 13 nm), and (d) 8.5\% (xy plane, z-axis range 11 to 13 nm). Green spheres represent FCC atoms, red spheres represent HCP atoms, and blue line represents 1/6$\langle112\rangle$ Shockley dislocation, purple line represents $1/6\langle110\rangle$ stair-rod dislocation. Gray spheres represent atoms at the \glspl{gb}.}
\label{fig:strain-stress}
\end{figure}

\begin{figure*}[!]
\centering
\includegraphics[width=1\linewidth]{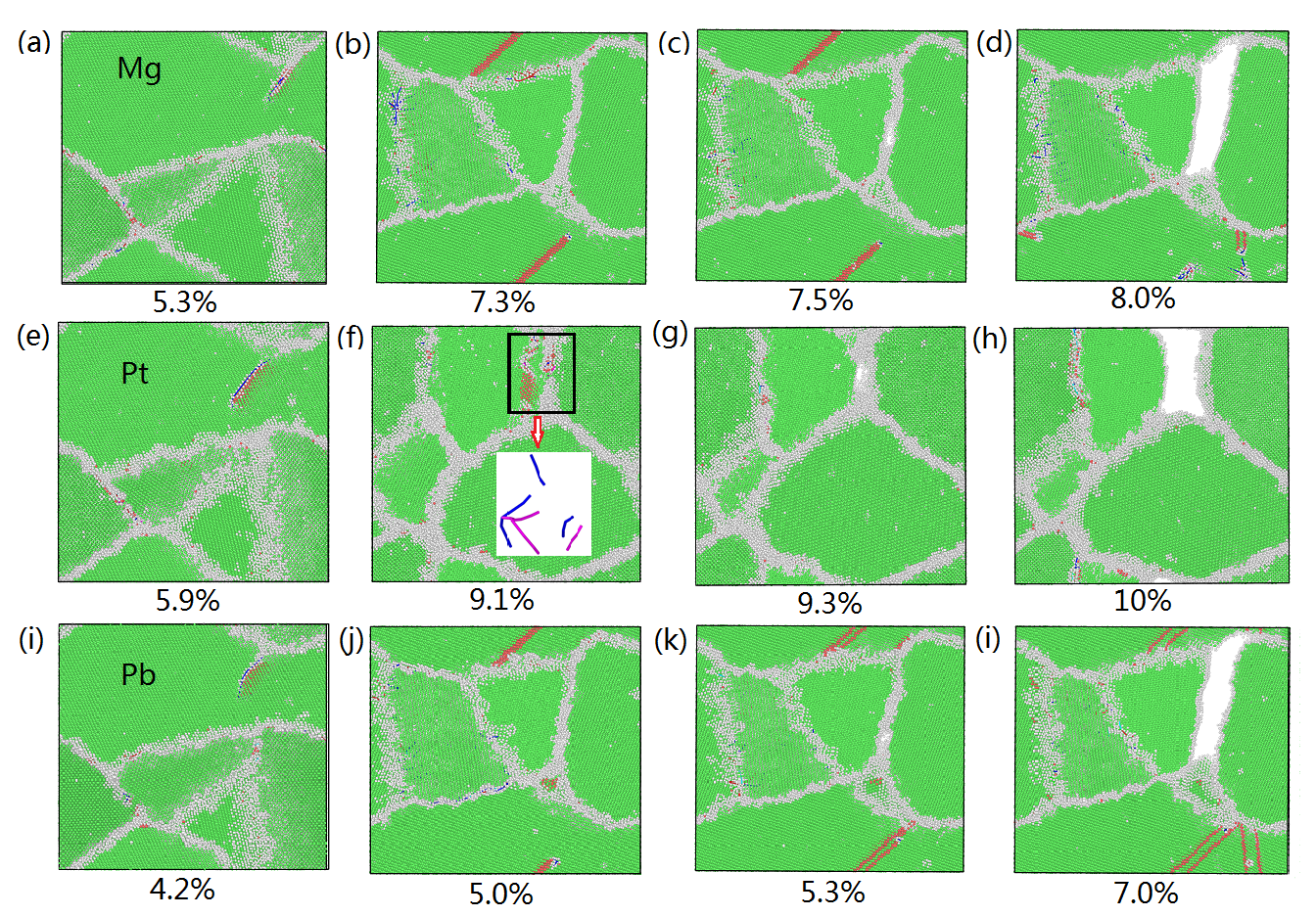}
\caption{Snapshots of 2-nm-thick slices for polycrystalline binary systems under different strains: (a-d) Mg, 5.3\% strain (yz plane, x-axis range 2 to 4 nm), the remaining strains (xy plane, z-axis range 12 to 14 nm);  (e-h) Pt, 5.9\% strain (yz plane, x-axis range 0.3 to 2.3 nm), the remaining strains (xy plane, z-axis range 4 to 6 nm); (i-l) Pb, 4.2\% strain (yz plane, x-axis range 2 to 4 nm), the remaining strains (xy plane, z-axis range 12 to 14 nm). Green spheres represent FCC atoms, red spheres represent HCP atoms, blue lines represent $1/6\langle112\rangle$ Shockley dislocations. In panel (f) inset, purple lines represent $1/6\langle110\rangle$ stair-rod dislocation. Pt segregation markedly enhances the cohesion of \glspl{gb}, with dislocation nucleation and transformation occurring earlier than crack nucleation, to accommodate the redistribution of stress during the deformation. The embrittling element does not exhibit this phenomenon.} 
\label{fig:strain}
\end{figure*}

\subsection{Impact of solute segregation on the strength and deformation mechanism of polycrystalline Al}

The relationship between tensile strength and strain of polycrystalline Al with various solutes is shown in Fig~\ref{fig:stress-strain-curves} (a). It is evident from the figure that Pt, Pd, Ta, Zr, Cr, V, Mo, Ti enhances the tensile strength of polycrystalline Al, whereas the remaining solutes reduce its strength. Strengthening solute (for Pt, Pd, and Zr) not only enhances the strength of polycrystalline Al but also enhances the maximum tensile strain. Since Ti Cr, V, Mo, and W do not segregate at \glspl{gb}, they will not be discussed further. 

Fig~\ref{fig:strain-stress} displays snapshots of pure polycrystalline Al under different deformations. Polycrystalline Al deformation goes through three stages. Initially, within the strain range of 0-4.9\%, the material exhibits elastic deformation. In the strain range of 4.9-8.3\%, dislocations and stacking faults are emitted from the \glspl{gb}. And at 8.3\% strain, cracks initiate at the \gls{gb} and then propagate along the \gls{gb}.

 Fig~\ref{fig:strain-stress} (a) shows the 1/6$\langle112\rangle$ Shockley dislocations and stacking faults in pure polycrystalline Al at a strain of 4.9\%. In the process of material deformation,  dislocations emitted from the \glspl{gb} play a decisive role\cite{Wu2006APL,Van2002PRB,Zhang2015MRE,Cui2023CMS}. Dislocation nucleation and propagation serve as the primary mechanisms for accommodating system stress. The slip and interplay of dislocations inside the grains facilitate stress redistribution throughout the plastic deformation phase.
 
When the strain reaches 8.3\%, cracks are formed, as shown in Fig~\ref{fig:strain-stress} (c), and the tensile stress reaches a maximum of 4.5 Gpa. In Fig~\ref{fig:strain-stress} (b), before crack formation, no defect formation and movement is observed in the crack initiation area. This indicates weak \gls{gb} cohesion, leading to intergranular cracking before the maximum stress value for defect nucleation. As strain increases, cracks propagate along the \gls{gb}, ultimately resulting in material fracture.

To further investigate the impact of solute segregation on the strength of polycrystalline Al, Pt and Pb were selected due to their significant influence on its strength. However, the influence of Mg solute remains contentious in DFT studies, with conflicting findings on whether Mg strengthens or embrittles \glspl{gb} \cite{Song1996AM,Liu1998AM,Zhang2012APL,Vsevolod2011SM}. Nevertheless, our analysis in Fig~\ref{fig:stress-strain-curves} reveals that Mg indeed embrittles polycrystalline Al, consistent with experimental observations\cite{Na1995SMM}. In Fig~\ref{fig:strain}, snapshots of Mg, Pt and Pb solute segregation in polycrystalline Al under different strains are presented.

 In Fig~\ref{fig:strain}(a), we observe that the strain (5.3\%) for Shockley dislocation nucleation is greater than that of pure polycrystalline Al (4.9\%), as illustrated in Fig~\ref{fig:strain-stress}. This can be attributed to the hindrance of dislocation nucleation and movement by Mg segregation at the \gls{gb} due to its large size. As shown in Fig~\ref{fig:stress-strain-curves} (b), the stress of Mg is greater than that of pure Al in the same strain (range 0-5.3\%); however, once dislocations are emitted, the stress curves overlap in the strain range of 5.3-7.5\%. Moreover, Mg may reduce the charge density at the \glspl{gb} \cite{ZHAO2018AM}, and Mg-Al bond weakens the \gls{gb} cohesion, leading to premature cracking of the \glspl{gb}. 

Unlike Mg, Pb promotes the formation of Shockley dislocation, as shown in Fig~\ref{fig:strain}(i). In MCMD simulations, Pd atoms segregate from the bulk to the \gls{gb}, leading to a decrease in the system's energy. Due to the significantly larger size of Pb compared to Al, the segregation of Pb at the loose positions of the \gls{gb} still causes expansion of both the \gls{gb} and the system, increasing local stress at the \gls{gb}. Consequently, this lowers the cohesion at the \gls{gb} and provides a driving force for the nucleation of dislocations. This size effect was also discussed at Ni and Ag \gls{gb}\cite{Lezzar2004AM}, consistent with our findings. In addition, Pb is also a well-known embrittling element\cite{Lejček2017PMS}, which significantly reduces the cohesion of \glspl{gb} and further promotes the nucleation of cracks at \glspl{gb}, as depicted in Fig~\ref{fig:strain}(k).

In contrast, Pt acts as a strengthening solute\cite{Lejček2017PMS}. When Pt segregates at Al \glspl{gb}, it increases the cohesion of \gls{gb}, thereby raising the stress required for dislocation nucleation and emission from the \glspl{gb}, as illustrated in Fig~\ref{fig:strain}(e). When the stress for crack nucleation at the \glspl{gb} exceeds the stress for dislocation and stacking fault nucleation and emission from the \gls{gb}, dislocation nucleation and movement occur prior to crack formation during the tensile process. As depicted in Fig~\ref{fig:strain}(f), before crack nucleation, the movement and transformation of Shockley dislocations (1/6$\langle112\rangle$ Shockley to 1/6$\langle110\rangle$ stair-rod dislocation) are observed in the crack area. The formation of the stair-rod dislocation marks the onset of plastic instability and indicates that cracks or holes will form\cite{Cui2023CMS,Ma2022JAC}.

\section{Summary and Conclusions}
In conclusion, we have successfully implemented a canonical-ensemble \gls{mcmd} algorithm in \textsc{gpumd}, demonstrating its accuracy and unprecedented efficiency. Leveraging this efficient \gls{mcmd} algorithm, we investigated the segregation behavior of 15 solutes in polycrystalline Al employing a general-purpose unified UNEP-v1 machine-learned potential for 16 elemental metals and their alloys \cite{Song2023ARX}. Our findings reveal distinct segregation patterns, with Ag, Au, Cu, Mg, Pb, Pd, Pt fully segregated at \glspl{gb}, while Ni, Ta, Mo and Zr partially segregated at GBs. Notably, Cr, V, W and Ti do not segregate at the \glspl{gb}, with Ti precipitating in the form of a BCC TiAl structure within the crystals. 

Furthermore, uniaxial tensile tests on 15 binary polycrystalline Al alloys demonstrate a significant reinforcement effect by Pt and Pd, whereas Pb induces substantial embrittlement. Through the analysis of microstructure, dislocation nucleation, and crack nucleation, it was found that during the deformation process of polycrystalline Al, \glspl{gb} play a pivotal role in determining its mechanical properties, and solute segregation significantly impacts the \glspl{gb} structure and cohesion, thereby affecting dislocation and crack nucleation, ultimately leading to strengthening or embrittlement of polycrystalline Al. 

Our efficient MCMD approach, coupled with unified neuroevolution potential in GPUMD, not only furnishes valuable tools for accurately and efficiently simulating solute segregation and chemical ordering in alloy systems, but also offers insights for future simulations and theoretical guidance for the design of polycrystalline alloy materials. We expect that our developed methods will also prove valuable for investigating a wide range of multi-component materials, including medium-entropy materials, high-entropy materials, and complex concentrated alloys.

\section{Acknowledgments}
We acknowledge support from the National Natural Science Foundation of China (NSFC) (No. 52071020).

\vspace{0.5cm}
\noindent{\textbf{Data availability:}}

The UNEP-v1 model is freely available at the Zenodo repository with following link: \url{ https://doi.org/10.5281/zenodo.10081677}.
The source code and documentation for \textsc{gpumd} are available
at \url{https://github.com/brucefan1983/GPUMD} and \url{https://gpumd.org}, respectively.

\vspace{0.5cm}
\noindent{\textbf{Declaration of competing interest:}}

The authors declare that they have no competing interests.

\appendix

\section{Input parameters for MCMD simulations in GPUMD}
\label{sec:run_in_MCMD}

\gls{mcmd} simulations with \gls{nep} models can be conducted using the \verb"gpumd" executable in the \textsc{gpumd} package \cite{Fan2017CPC}. The parameters controlling the simulations are specified in the \verb"run.in" input file. Below is an example of the contents of the \verb"run.in" input file for one system. The keyword \verb"mc" invokes the \gls{mc} trials during the \gls{md} simulation. The parameter \verb"canonical" indicates that the \gls{mc} ensemble is canonical, conserving the number of atoms for each species. The parameters \verb"100 100" mean performing 100 \gls{mc} trails after every 100 \gls{md} stesp. The parameters \verb"300 300" denote maintaining the temperature in the \gls{mc} trials at 300 K.

\begin{verbatim}
# setup
potential nep.txt
velocity  300

# MCMD simulation
ensemble     npt_mttk temp 300 300 iso 0 0
time_step    1
mc           canonical 100 100 300 300
dump_thermo  100
dump_exyz    50000
dump_restart 10000
run          5000000
\end{verbatim} 

\section{Inputs for tensile loading simulations in GPUMD}
\label{sec:run_in_tensile}

Tensile loading simulations with \gls{nep} models can also be performed using the \verb"gpumd" executable in the \textsc{gpumd} package \cite{Fan2017CPC}. The relevant contents of the \verb"run.in" input file are provided below. Tensile loading is initiated by the keyword \verb"deform", with a deformation speed of $2\times 10^{-5}$ \AA/fs, corresponding to an engineering strain rate of $10^{8}$ s$^{-1}$ for a system with a linear size of 200 \AA. Due to an implementation restriction in \textsc{gpumd}, the \verb"deform" keyword can only be effective by using an $NpT$ ensemble. Therefore, we used a very large relaxation time for the barostat to effectively convert the $NpT$ ensemble into the $NVT$ ensemble intended for use.

\begin{verbatim}
# tensile loading simulation
ensemble     npt_scr 300 300 100 
             0 0 0 100 100 100 1e10
time_step    1
deform       0.00002 0 1 0
dump_thermo  100
dump_exyz    1000
dump_restart 10000
run          3000000
\end{verbatim}


%

\end{document}